\documentclass[fleqn,twoside,twocolumn,nofootinbib,showkeys]{revtex4} 
\usepackage[nocpr]{ujp} 

\begin{document}
\title[Resonant Enhancement of Molecular Excitation Intensity]
{RESONANT ENHANCEMENT\\ OF MOLECULAR EXCITATION INTENSITY\\ IN
INELASTIC ELECTRON SCATTERING\\ SPECTRUM OWING TO INTERACTION\\ WITH
PLASMONS IN METALLIC NANOSHELL}
\author{I.Yu. Goliney}
\affiliation{Institute for Nuclear Research, Nat. Acad. of Sci. of Ukraine}
\address{47, Nauky Ave., Kyiv 03680, Ukraine}
\email{igoliney@kinr.kiev.ua}
\author{Ye.V.~Onykienko}%
\affiliation{Institute for Nuclear Research, Nat. Acad. of Sci. of Ukraine}%
\address{47, Nauky Ave., Kyiv 03680, Ukraine}%
\email{igoliney@kinr.kiev.ua} \udk{???}  \razd{}

\autorcol{I.Yu.\hspace*{0.7mm}Goliney,
Ye.V.\hspace*{0.7mm}Onykienko}

\setcounter{page}{922}%

\begin{abstract}
A quantum-mechanical model to calculate the electron energy-loss
spectra (EELS) for the system of a closely located metallic
nanoshell and a molecule has been developed.\,\,At the resonance
between the molecular excitation and plasmon modes in the nanoshell,
which can be provided by a proper choice of the ratio of the inner
and outer nanoshell radii, the cross-section of inelastic electron
scattering at the molecular excitation energy is shown to grow
significantly, because the molecular transition borrows the
oscillator strength from a plasmon.\,\,The enhancement of the
inelastic electron scattering by the molecule makes it possible to
observe molecular transitions with an electron microscope.\,\,The
dependences of the EEL spectra on the relative arrangement of the
molecule and the nanoshell, the ratio between the inner and outer
nanoshell radii, and the scattering angle are plotted and analyzed.
\end{abstract}
 \maketitle

\section{Introduction}

In the last years, a relatively new domain of researches,
plasmonics, develops intensively \cite{Maier}.\,\,It uses the
ability of collective electron oscillations in metals to enormously
strengthen electric fields \cite{Klimov} for studying various
physical systems and developing new methods of researches in science
\cite{Hirakawa}, engineering \cite{Atwater}, and medicine
\cite{Gobin}.

The plasmon resonance is studied using the methods based on the
interaction of plasmons with light, electrons, or atoms.\,\,One of
those methods is the electron energy-loss spectroscopy
(EELS).\,\,The high resolution of an electron microscope (down to
1~\AA ) allows the effect of plasmon resonance on nano-dimensional
objects to be observed with a better spatial resolution than other
spectroscopic methods permit.\,\,In the recent years, the EELS was
used to study nanoparticles of various shapes \cite{Nelayah}, the
interaction between nanoparticles \cite{Ugarte}, thin films
\cite{Chen}, and composite nano-materials \cite{McComb}.\,\,The
authors of work \cite{Nicoletti1} studied plasmon excitations in
silver nanorods and showed that the plasmon excitations are
quantized into resonance modes, with the intensity of mode maxima
changing along the nanorod, and the plasmon wavelength being minimum
at the nanorod ends.\,\,In work \cite{Nicoletti2} on the basis of
electron tomography, the method of localized plasmon mode imaging
was developed using a three-dimensional nano-sized cube as an
example, and a capability to single-out every mode from the total
output signal with regard for the substrate and the shape of a
nano-dimensional object was demonstrated.

This work aims at studying the possibility to observe the excitation
of an organic molecule in EEL spectra owing to the resonant
interaction between the molecule and localized plasmons in a
nanoparticle or a nanoshell.\,\,The oscillator strength of the
molecule in the excited state can considerably grow at its resonant
mixing with plasmon excitations in the nanoparticle due to a large
dipole moment of localized plasmons.\,\,This enhancement is
especially pronounced in the surface enhancement of Raman spectra
\cite{SERS} and, to some less extent, in fluorescence spectra
\cite{FLUOR}.\,\,A similar effect should expectedly take place for
the EEL spectra as well, which would enable separate molecules
adsorbed on the nanoparticle surface to be observed with an electron
microscope.

For the theoretical calculation of EEL spectra, a quantum-mechanical
approach developed in works \cite{gol1,gol2} is applied.\,\,This
procedure includes the calculation of the classical plasmon spectrum
for a metallic nanoparticle, the quantization of normal modes, the
calculation of the spectrum of the composite system taking the
plasmon interaction and the molecule excitation into account, and
the calculation of the cross-sections of inelastic electron
scattering with the excitation of the obtained states.

The mixing of localized plasmon states with mole\-cu\-lar ones has a
resonant character.\,\,At the same ti\-me, the properties of
localized plasmons depend on na\-no\-par\-tic\-le's shape
\cite{Ciraci,Kern} and size \cite{Scholl}.\,\,This
cir\-cum\-stan\-ce makes it possible to reach the effective
interaction between plas\-mons and the molecule by choo\-sing the
corresponding shape of a na\-no\-par\-tic\-le.\,\,The inelastic
electron scattering at a spherical nanoparticle was considered in
work \cite{NAUKMA}.\,\,In the present work, we analyze the more
complicated case of a spherical nanoshell.

\section{Model of the System}

The system consisting of a molecule near a spherical metallic
nanoshell is considered (Fig.\,\,1).\,\,The spectrum of molecular
excitations is simulated as a two-level system with the transition
moment $\mathbf{d}$.\,\,The distance between the molecule and the
nanoparticle surface equals $L$.\,\,The outer radius of the
nanoshell is $R_{2}$, the inner one is $R_{1}$.\,\,The nanoshell
radius $R_{2}$ is supposed to be small in comparison with the length
of electromagnetic waves, whose frequency corresponds to the plasmon
frequency, $R_{2}\ll c/\omega_{p}$, where $c$ is the speed of light,
and $\omega_{p}$ is the plasma frequency.\,\,The molecule can be
oriented arbitrarily with respect to the nanoparticle and the
high-energy electron beam propagation direction.\,\,We suppose that
an electron beam with wavevector $\mathbf{k}$ falls on the system
concerned and is scattered into a state with the wavevector
$\mathbf{k}^{\prime}$.

The spectrum of nanoshell plasmon oscillations has two modes, which
correspond roughly to oscillations on the outer and inner surfaces.
Therefore, the first problem consists in finding the frequencies of
those modes and their dependences on the ratio between the inner and
outer radii. Below, it will be demonstrated that, by choosing this
ratio, it is possible to reach a resonance between the excitation
energies of the molecule and a localized plasmon.

\begin{figure}%
\vskip1mm
\includegraphics[width=\column]{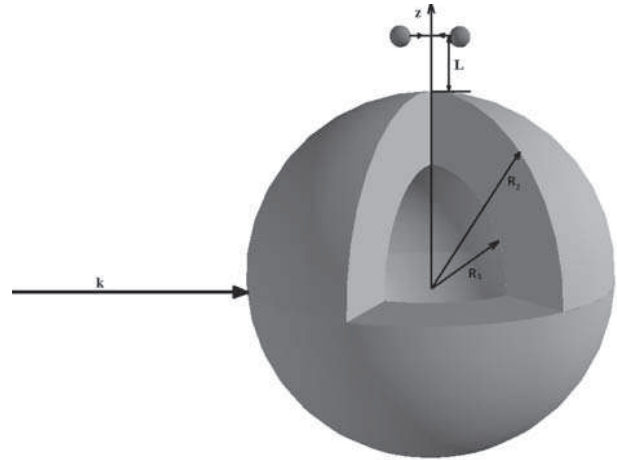}
\vskip-3mm\caption{Scheme of the electron scattering by the system
consisting of a spherical nanoshell with inner radius $R_{1}$ and
outer radius $R_{2}$, and the molecule-dipole at a distance $L$
}\label{ris:shell}
\end{figure}

The next step in the calculation procedure is the quantization of
plasmon oscillations and the determination of the mixed-mode
spectrum  taking into account the interaction between the molecule
modeled as a two-level system and localized plasmons.\,\,The
probability of the electron scattering with the excitation of those
mixed modes is calculated in the Born approximation.

\section{Surface Plasmons\\ on a Spherical Metallic Nanoshell}

The nanoshell is considered in the quasistatic approximation,
i.e.\,\,the nanoparticle radius is supposed to be much smaller than
the wavelength, $R\ll c/\omega$.\,\,This approximation allows us to
characterize the electromagnetic field by the electrostatic
potential $\phi$.

To find the spectrum of localized plasmons in the nanoshell, we have
to consider the electromagnetic field in three regions: the
dielectric core, metallic shell, and external region.\,\,The field
in the metallic shell interacts with free electrons.\,\,We will
consider the electron subsystem of the metal, by using the Drude
model.

The polarization $\mathbf{P}$ of the electron subsystem in the
electrostatic potential~-- as a consequence of Newton's equation of
motion for free electrons, $m\ddot{\mathbf{r}}=-e\mathbf{E}=$
$=e\nabla\phi$~-- is described by the\vspace*{-1mm}
equation\begin{equation}
\ddot{\mathbf{P}}=-en\ddot{\mathbf{r}}=-\frac{ne^{2}}{m}\nabla\phi,
\label{Poleq}
\end{equation}\vspace*{-5mm}

\noindent where $m$ is the electron mass, $e$ the electron charge,
and $n$ the electron concentration. This equation has to be solved
together with Maxwell's equation for the electric field,
\begin{equation}
\nabla \cdot \mathbf{D} = \varepsilon_{0} \nabla^2\phi  + 4\pi
\nabla \cdot \mathbf{P} = 0, \label{Max4}
\end{equation}
where $\varepsilon_{0}$ is the contribution of all factors, except for free
electrons, to the dielectric permittivity.

It is convenient to seek the polarization in the form
$4\pi\mathbf{P}=-\nabla\psi$, where $\nabla^{2}\psi=0$.\,\,Then the
field equation (\ref{Max4}) reads
\begin{equation}
\nabla^{2}\phi=0,   \label{laplace}
\end{equation}
and Eq.~(\ref{Poleq}) transforms into
\begin{equation}
\ddot{\psi}=\omega_{p}^{2}\phi,   \label{Poleq2}
\end{equation}
where $\omega_{p}=\sqrt{\frac{4\pi e^{2}n}{m}}$ is the plasma
fre\-qu\-en\-cy.\,\,Equa\-tions (\ref{laplace}) and (\ref{Poleq2})
have to be solved together with the continuity conditions for the
potential $\phi$ and the normal component of the electric induction
vector,
\begin{equation}
\mathbf{D_{n}}=-\varepsilon_{0}\nabla\phi-\nabla\psi.
\end{equation}

The fields in three regions~-- the dielectric core of the nanoshell,
the metallic layer, and the environment~-- are sought in the
following general forms:
\begin{equation}\label{field1}
\phi^{(1)} = \sum_{lm} a_{lm}^{(1)} r^l Y_{lm}(\theta, \varphi),
\end{equation}\vspace*{-5mm}
\begin{equation}\label{field2}
\left\{
\begin{aligned}
\!\!\phi^{(2)} = \sum_{lm} ( a_{lm}^{(2)} r^l + b_{lm}^{(2)} r^{-l -1} ) Y_{lm}(\theta, \phi),\\
\!\!\psi^{(2)} = \sum_{lm} ( c_{lm}^{(2)} r^l + d_{lm}^{(2)} r^{-l -1} ) Y_{lm}(\theta, \phi),\\
\end{aligned}
\right.
\end{equation}\vspace*{-5mm}
\begin{equation}\label{field3}
\phi^{(3)} = \sum_{lm} a_l^{(3)} r^{-l-1} Y_{lm}(\theta, \varphi).
\end{equation}

Applying the continuity conditions for the potential and the normal
component of the electric induction vector, we obtain a system of equations
for the coefficients $a_{lm}$, $b_{lm}$, $c_{lm}$, and
$d_{lm}$:
\begin{equation}
\left\{
\begin{array}{l}
 a_{lm}^{(1)} = a_{lm}^{(2)} + b_{lm}^{(2)} R_1^{-2l -1},\\[2mm]
 a_{lm}^{(2)} + b_{lm}^{(2)} R_2^{-2l -1} = b_{lm}^{(3)} R_2^{-2l -1} ,
 \\[2mm]
-\varepsilon^{(1)} l a_{lm}^{(1)} = \mathbf{D_n}^{(2)},\\[2mm]
-\varepsilon^{(3)} \frac{l+1}{R_2^{2l+1}}  b_{lm}^{(3)} = \mathbf{D_n}^{(2)},\\
\end{array}
\right.
\end{equation}
where
$D_{\mathbf{n}}^{(2)}=\varepsilon_{0}\frac{l+1}{R_{1}^{2l+1}}b_{lm}^{(2)}-\varepsilon_{0}la_{lm}^{(2)}+\frac{l+1}{R_{1}^{2l+1}}%
d_{lm}^{(2)}-lc_{lm}^{(2)}$ is the normal component of the electric
induction vector in the metallic layer.

The solution of this system makes it possible to relate all unknown
variables to the coefficients $c_{lm}^{(2)}$ and $d_{lm}^{(2)}$ that
express the electromagnetic field in the metallic
layer.\,\,Substituting those solutions into Eq.~(\ref{Poleq2}), we
obtain a system of equations for the plasmon modes,\vspace*{-1mm}
\begin{equation} \label{oscil1}
 \left\{\!
\begin{aligned}
\ddot{c}_{lm}^{(2)} - \omega_p^2 A_{lm}^{11} {c}_{lm}^{(2)} - \omega_p^2  A_{lm}^{12} {d}_{lm}^{(2)} = 0, \\
\ddot{d}_{lm}^{(2)} - \omega_p^2 A_{lm}^{21} {c}_{lm}^{(2)} -
\omega_p^2 A_{lm}^{22} {d}_{lm}^{(2)} = 0.
\end{aligned}
\right.
\end{equation}
The coefficients $A_{lm}$ are given in Appendix.

System (\ref{oscil1}) is a system of equations for coupled
oscillators.\,\,The corresponding equations for two normal modes for
every $l$ and $m$ look like\vspace*{-1mm}
\begin{equation}
\ddot{q}_{lm}^{\pm}+\left( \omega_{l}^{\pm}\right) ^{2}q_{lm}^{\pm}=0.
\end{equation}
The normal mode frequencies are determined by the
formula\vspace*{-2mm}
\[
(\omega_l^{(\pm)}/ \omega_p)^2 = - \frac{A_{lm}^{11} +
A_{lm}^{22}}{2} \,\pm \]\vspace*{-6mm}
\begin{equation}
\pm\, \sqrt{\frac{(A_{lm}^{11} - A_{lm}^{22})^2}{4} +
A_{lm}^{21}A_{lm}^{12}}.
\end{equation}
Accordingly, the coefficients $c_{lm}^{(2)}$ and $d_{lm}^{(2)}$ in
Eqs.\,(\ref{oscil1}) are expressed in terms of the coefficients
$q_{lm}^{\pm}$ as follows:\vspace*{-1mm}
\begin{equation}\label{clm}
c_{lm}^{(2)} = q_{lm}^+ + q_{lm}^-,
\end{equation}\vspace*{-5mm}
\[\label{dlm}
d_{lm}^{(2)} = -\frac{ (\omega_l^{+})^2 + \omega_p^2
A^{11}_{lm}}{\omega_p^2 A^{12}_{lm}} q_{lm}^+ \,- \]\vspace*{-5mm}
\begin{equation}
- \, \frac{ (\omega_l^{-})^2
 + \omega_p^2 A_{lm}^{11}}{\omega_p^2 A_{lm}^{12}}q_{lm}^-.
\end{equation}

In Fig.~2, the oscillation frequencies versus the ratio between the
outer and inner nanoshell radii are shown for two dipole modes
($l=1$).\,\,If the metallic layer thickness is large, those modes
can be regarded as plasmons localized on the inner and outer
surfaces of the shell, and their frequencies are equal to the
frequencies of localized plasmons in a solid particle and the pore
in a metal, respectively.\,\,As the metallic layer becomes thinner,
the interaction between oscillations on two surfaces gets stronger,
which gives rise to considerable variations in the spectrum.\,\,This
is especially actual for the lower mode, which corresponds to the
outer surface.\,\,For a silver nanoshell, the choice of the ratio
between the radii allows one to reduce the plasmon frequency from
the ultra-violet spectral range to the visible and even infra-red
one if the metallic layer is very thin.\,\,Hence, we obtain a
possibility to select nanoshells, the plasmon oscillations in which
will be in resonance with the molecular transition
frequency.

\subsection{Quantization of the nanoshell field}

While determining the intensity of the interaction between plasmon
oscillations in the nanoshell and other particles~-- atoms,
molecules, or high-energy electrons~-- we used the quantization
approach for the electromagnetic field of plasmons, which was
developed and applied in works \cite{gol1,
gol2,sugver,golchph}.\,\,The quantization procedure begins with the
determination of the proper action functional: by varying its
variables, we can obtain the equations of motion.\,\,In the case of
a nanoshell, the action functional must be a start point for the
derivation of the equation for the electromagnetic field, the
equation of motion for the electron polarization in the metallic
layer, and the boundary conditions at the inner and outer
surfaces.\,\,The corresponding action looks like\vspace*{-2mm}
\begin{equation}
S=\int\limits_{0}^{t}(\mathcal{L}_{\rm core}+\mathcal{L}_{\rm metal}+\mathcal{L}%
_{\rm outer})dt,
\end{equation}
where\vspace*{-3mm}
\begin{equation} \mathcal{L}_{\rm
core}=\frac{1}{4\pi}\int\limits_{V_{\rm core}}\frac{\varepsilon
^{(1)}}{2}(\nabla\phi)^{2}dV,
\end{equation}\vspace*{-5mm}
\[
\mathcal{L}_{\rm metal} = \frac{1}{4 \pi} \int\limits_{V_m} \Biggl[
\frac{\varepsilon_{0}}{2} (\nabla \phi)^2 - 4 \pi \mathbf{P}\nabla
\, \phi \,+  \]\vspace*{-5mm}
\begin{equation}
  +\left(\!\frac{4\pi}{\omega_p}\!\right)^{\!\!2}
\frac{1}{2}\, \dot{\mathbf{P}}^2  \Biggr] dV,
\end{equation}\vspace*{-5mm}
\begin{equation}
\mathcal{L}_{\rm outer} = \frac{1}{4 \pi} \int\limits_{V_{\rm core}}
\frac{\varepsilon^{(3)}}{2} (\nabla \phi)^2 dV.
\end{equation}

\begin{figure}%
\vskip1mm
\includegraphics[width=\column]{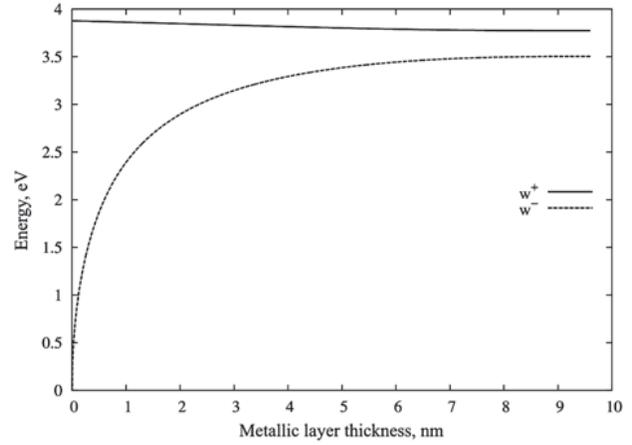}
\vskip-3mm\caption{Dependences of the plasma oscillation energy on
the silver nanoshell thickness.\,\,The outer radius equals 10~nm,
and the inner one changes from 0.5 to 9.5~nm
}\label{ris:frequency}\vspace*{-2mm}
\end{figure}

\noindent Substituting the corresponding expressions for the fields,
Eqs.\,(\ref{field1})--(\ref{field3}), and carrying out a cumbersome
integration, we reduce the Lagrangian density to the classical
oscillator form,
\[
\mathcal{L} = \frac{1}{2} \sum_{lm} \biggl\{\! \left[
\left(\dot{Q}_{lm}^+ \right)^{\!2} - \left(\omega_l^+ \right)^2
\left( Q_{lm}^+ \right)^2 \right] +  \]\vspace*{-6mm}
\begin{equation}
  + \left[
\left(\dot{Q}_{lm}^- \right)^{\!2} - \left(\omega_l^+ \right)^2
\left( Q_{lm}^- \right)^2 \right]\!\biggr \},
\end{equation}
where the variables $Q_{lm}^{-}$ and $Q_{lm}^{+}$ are proportional
to the variables $Q_{lm}^{\pm}=\sqrt{B_{lm}^{\pm}}q_{lm}^{\pm}$
introduced above (the $B_{lm}$-values are given in Appendix).

The transition to quantum-mechanical operators is carried out, by
following the canonical quantization procedure of oscillator
variables and equating the quantum energy to
$\hbar\omega_{l}^{\pm}$.\,\,As a result of the quantization, we can
write the electric potential of plasma oscillations in terms of the
plasmon creation, $\hat{Q}_{lm}^{\mu\dagger}$, and annihilation,
$\hat{Q}_{lm}^{\mu}$, operators,
\begin{equation}\label{qfield1}
\hat{\phi^{(1)}} = \sum_{lm \mu}  \sqrt{\frac{\hbar}{2\omega_l^\mu
B_l^\mu}} N_l^\mu r^l (\hat{Q}_{lm}^{\mu\dagger} +
\hat{Q}_{lm}^{\mu}) Y_{lm}(\theta, \varphi),
\end{equation}\vspace*{-7mm}
\[ \hat{\phi^{(2)}} = \sum_{lm\mu}
  \sqrt{\frac{\hbar}{2\omega_l^\mu B_l^\mu}} \left(\!
  F_l^\mu\frac{1}{r^{l+1}}+\left(\!\frac{\omega
    _l^\mu}{\omega_p}\!\right)^{\!\!2}r^l\!\right)\! \times \]\vspace*{-6mm}
\begin{equation}\label{qfield2}
     \times\,  (\hat{Q}_{lm}^{\mu\dagger} +
  \hat{Q}_{lm}^{\mu})Y_{lm}(\theta, \varphi),
\end{equation}\vspace*{-7mm}
\[
\hat{\phi^{(3)}} = \sum_{lm\mu} \sqrt{\frac{\hbar}{2\omega_l^\mu
B_l^\mu}} K_l^\mu \,\times
\]\vspace*{-6mm}
\begin{equation}\label{qfield3}
 \times\,\frac{1}{r^{l+1}} (\hat{Q}_{lm}^{\mu\dagger} +
\hat{Q}_{lm}^{\mu}) Y_{lm}(\theta, \varphi),
\end{equation}
where the subscript $\mu$ acquires the values
\textquotedblleft$+$\textquotedblright\ and
\textquotedblleft$-$\textquotedblright, and the formulas for
$N_{l}$, $F_{l}$, and $K_{l}$ are given in Appendix.\,\,Below, when
summing up over $l$, only the first term of the series will be taken
into consideration, because the higher terms give an insignificant
contribution to the field expression, i.e.\,\,we are limited the
dipole approximation ($l=1$).

\subsection{Mixed excitation spectrum\\ for the molecule and the nanoshell}

\label{system}

The interaction between electrons in the molecule and the electron
subsystem in the nanoparticle results in the mixing of quantum
states in both subsystems of our system.\,\,This mixing can be
substantial if it has a resonant character, i.e.\,\,if the energy of
the excited state of the molecule is close to the energy of a
plasmon in the nanoparticle.\,\,We consider the molecule as a
two-level system with the dipole moment $\mathbf{d}$ (Fig.~1) and
the creation and annihilation operators $\hat{c}^{\dagger}$ and
$\hat{c}$, respectively.

The total Hamiltonian of the system \textquotedblleft molecule +
nanoparticle\textquotedblright, which involves the interaction
between them in the dipole approximation, looks like\vspace*{-1mm}
\begin{equation}
H=\sum\limits_{1m\mu}(\hbar\omega_{1}^{\mu}\hat{Q}_{1m}^{\mu\dagger}\hat
{Q}_{1m}^{\mu})+E_{0}\hat{c}^{\dagger}\hat{c}-\mathbf{d}\nabla\phi
(\mathbf{r})(\hat{c}^{\dagger}+\hat{c}).   \label{hmix}
\end{equation}
The interaction between the molecule and the nanoshell is described
by the last term in this Hamiltonian, in which the operator of
nanoparticle electrostatic potential is taken at the point
 $\mathbf{r}$, where
the molecule is located.\,\,The diagonalization of Hamiltonian
(\ref{hmix}) gives us the spectrum and the wave functions of the
system.

It is convenient to change from the operators
$\hat{Q}_{lm}^{\mu\dagger}$ and $\hat{Q}_{lm}^{\mu}$ to the
operators $Q_{p}^{\mu}$ and $Q_{p}^{\mu\dagger}$, where the
subscript $p$ runs through the values $\left( x,y,z\right) $
according to the relations\vspace*{-1mm}
\[\hat{Q}_x^{\mu\dagger}\!=
(\hat{Q}_{1-1}^{\mu\dagger} -
\hat{Q}_{11}^{\mu\dagger})/\sqrt{2},\]\vspace*{-8mm}
\[\hat{Q}_y^{\mu\dagger}\!=\imath
(\hat{Q}_{1-1}^{\mu\dagger}+\hat{Q}_{11}^{\mu\dagger})/\sqrt{2},\]\vspace*{-8mm}
\[
 \hat{Q}_z^{\mu\dagger}=\hat{Q}^{\mu\dagger}_{10}.
\]

The eigenvalues of the Hamiltonian~-- there are  7 of them~-- give
the energy spectrum of the system. The spectrum of plasmons in the
nanoshell is triple-degenerate with respect to the magnetic quantum
number $m$ of the molecule.\,\,However, the presence of the molecule
results in a partial lift of this degeneration.\,\,The obtained
spectrum contains two degenerate levels of surface plasmons for each
of the energies $\hbar \omega_{1}^{+}$ and
$\hbar\omega_{1}^{-}$.\,\,The third level shifts as a result of the
interaction with the molecule.\,\,The split of surface plasmon
levels and a shift of the molecular excitation energy depend on the
distance between the dipole and the particle and on the orientation
of the dipole moment of a molecular transition.

The creation operators for the mixed states are determined in terms of
characteristic vectors of the whole system and are given by the
relations
\begin{equation}
\hat{\psi}_{j}^{\dagger}=\sum\limits_{p=1}^{3}S_{jp}\hat{Q}_{p}^{+\dagger
}+\sum\limits_{p=4}^{6}S_{jp}\hat{Q}_{p}^{-\dagger}+S_{j7}\hat{c}^{\dagger},
\end{equation}
where $S_{jp}$ is a unitary matrix determined while diagonalizing
the Hamiltonian (in the course of further calculations, the
Hamiltonian was diagonalized numerically).\,\,Accordingly, taking
into account that $S_{jp}$ is a real-valued matrix, the creation
operators $\hat{Q}_{p}^{\mu\dagger}$ required for the further
calculations can be expresses in terms of the new operators of mixed
excitations, $\hat{\psi}_{j}^{\dagger}$, as follows:
\begin{equation}
\hat{Q}_{p}^{+\dagger}=\sum\limits_{j=1}^{7}S_{pj}\hat{\psi}_{j}^{\dagger
},\quad\hat{Q}_{s}^{-\dagger}=\sum\limits_{j=1}^{7}S_{sj}\hat{\psi}%
_{j}^{\dagger},   \label{Q}
\end{equation}
where $p=\left\{ 1,2,3\right\} $ and $s=\left\{ 4,5,6\right\}
$.\,\,For the creation operator of a molecular excitation
$\hat{c}^{\dagger}$, we have
\begin{equation}
\hat{c}^{\dagger}=\sum_{j=1}^{7}S_{7j}\hat{\psi}_{j}^{\dagger}.   \label{c}
\end{equation}
Similar expressions can be obtained for the annihilation operators
$\hat{Q}_{p}^{+}$, $\hat{Q}_{s}^{-}$, and $\hat{c}$.

\section{Calculation of Electron Scattering Spectra}

Let the electron, which is regarded as a quantum-mechanical particle
with the initial energy $E$, be scattered by the system
\textquotedblleft molecule--nanoshell\textquotedblright.\,\,The wave
function of the incident electron is a plane wave with the wave
vector $\mathbf{k}$.\,\,A detector registers electrons with the wave
vector $\mathbf{k}^{\prime}$.\,\,The general scattering setup is
illustrated in Fig.~3.

The general form of the Hamiltonian for an electron interacting with the
examined system looks like
\begin{equation}
\hat{H}=-\frac{\hbar^{2}\Delta}{2m_{e}}+\hat{H}_{S}+\hat{H}_{\rm
int},
\end{equation}
where the first term is the operator of the electron kinetic energy,
$\hat {H}_{S}$ is the Hamiltonian of the system, and $\hat{H}_{\rm
int}$ the operator of interaction between the electron and the
system.

The Hamiltonian of the system $\hat{H}_{S}$ can be written in the
form of a series expansion in the normal oscillation modes of the
system.\,\,Then, the operator of interaction between the system and
the external electron, $\hat {H}_{S}=-e\Phi(\mathbf{r})$, contains
terms proportional to the creation and annihilation operators of
normal modes.\,\,The interaction intensity is calculated as the
electrostatic potential of the excitation at the point, where the
electron is located.

The effective electron scattering cross-section per unit solid angle can be
determined from the general relation \cite{Landau}
\begin{equation}
d\sigma=|f(\mathbf{k},\mathbf{k^{\prime}})|^{2}d\Omega,
\end{equation}
where $f(\mathbf{k},\mathbf{k^{\prime}})$ is the electron scattering
amplitude.\,\,We consider electrons to be fast enough to treat the
scattering field $\Phi$ as a perturbation.\,\,Under those
conditions, the formula for the scattering amplitude in the
Born--Oppenheimer approximation looks like
\begin{equation}
f(k,k^{\prime})=-\frac{m_{e}}{2\pi\hbar^{2}}V_{\mathbf{Q}n}=-\frac{m_{e}}{%
2\pi\hbar^{2}}\int e^{i\mathbf{Q}\mathbf{r}}\Phi(\mathbf{r})d^{3}r,
\label{matrix}
\end{equation}
where $V_{\mathbf{Q}n}$ is the matrix element for the electron transition
from the state $k$ into the state $k^{\prime}$ with the excitation of the
$n$-th mode in the system, and
$\mathbf{Q}\,{=}\,\mathbf{k}\,{-}\,\mathbf{k^{\prime}}$ is the transferred momentum, as is shown in Fig.~3.

In order to calculate the double differential scattering cross-section
defined by the expression
\begin{equation}
\frac{d\sigma}{d\Omega
dE}=\frac{m^{2}}{4\pi^{2}\hbar^{4}}V_{\mathbf{Q}n}^{2}\delta(\Delta E-E_{l}),   \label{doublecross}
\end{equation}
where $\Delta E=(k^{2}-k^{\prime2})/2m$ is the energy lost by the
electron, and $E_{l}$ are the resonance energies, it is necessary to
find the transition matrix element $V_{\mathbf{Q}n}$.\,\,In the
following sections, we will separately consider the cases of the
electron scattering by the nanoshell, the molecule, and the system
composed of the nanoshell and the molecule.

\begin{figure}%
\vskip1mm
\includegraphics[width=\column]{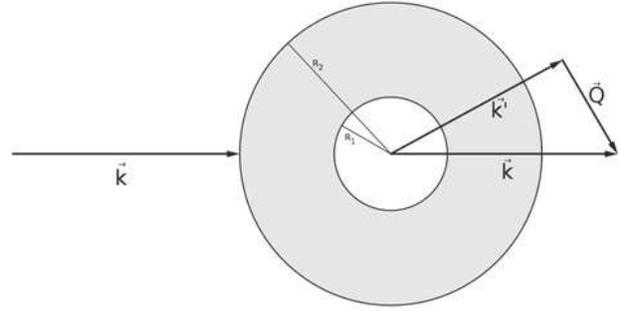}
\vskip-3mm\caption{Scheme of the electron scattering by a spherical
nanoshell with inner radius $R_{1}$ and outer radius
$R_{2}$}\label{ris:shell2}
\end{figure}

\subsection{Inelastic electron\\ scattering by the nanoshell}

\label{nanoshell0}

For the electron scattered by a conducting nanoshell, the interaction
Hamiltonian looks like
\begin{equation}
H_{\rm int}=-e\phi(\mathbf{r}),
\end{equation}
where $\phi(\mathbf{r})$ is the field determined by formulas
(\ref{qfield1}), (\ref{qfield2}), and (\ref{qfield3}) for three
regions: the core, metallic layer, and external space,
respectively.\,\,Only three modes are excited in the dipole
approximation: with $l=1$ and $m=0,\pm1$.\,\,It is convenient to
calculate the matrix element in a coordinate system with the
applicate axis directed along the vector $\mathbf{Q}$ and choose the
same coordinate frame for normal modes.\,\,In this frame, only the
mode with $m=0$ is excited.

In the spherical coordinate system, the matrix element of the electron
transition reads
\begin{equation}
V_{Q,1,0}^{(\rm shell)}=-e\iiint \sin {\theta \,}r^{2}e^{iQr\cos
{\theta }}\phi (r)drd\theta d\varphi.
\end{equation}It is convenient to use the formula
\begin{equation}
e^{iQr\cos {\theta }}=\sum_{l=0}^{\infty }(2l+1)i^{l}j_{l}(Qr)P_{l}(\cos
{\theta }).  \label{exp_expand}
\end{equation}%
Hence, it remains only to integrate over the radial variable, with
the corresponding interval being divided into three regions: the
nanoshell core, metallic layer, and region outside the
nanoshell.\,\,Those integrals are reduced to the following ones to
within coefficients: $I_{1}=\int_{0}^{R}\chi ^{3}j_{1}(\alpha \chi
)d\chi $, $I_{2}=\int_{R}^{1}\chi ^{3}j_{1}(\alpha \chi )d\chi $,
$I_{3}=\int_{R}^{1}j_{1}(\alpha \chi )d\chi $, and $I_{4}=
\int_{1}^{\infty }j_{1}(\alpha \chi )d\chi $ with the dimensionless
variables $\chi =r/R_{2}$, $\alpha =QR_{2}$, and $R=R_{1}/R_{2}
$.\,\,The matrix elements $V_{Q,1,0}^{(\rm shell)}$ can be the
calculated with the help of the formula
\[
V_{Q, 1,0}^{(\rm shell)} = i\sqrt{\frac{27\pi\hbar e^2}{2
\omega_1^\mu B_1^\mu}}\biggl\{ R_2^4N_1^\mu I_1+R_2^4\left(\!
\frac{\omega _1^\mu}{\omega _p}\!\right)^{\!\!2} I_3\, +
\]\vspace*{-5mm}
\begin{equation}\label{nanoshell}
  +\,R_2F_1^\mu
I_2+R_2K_1^\mu I_4\biggr\}\! ,
\end{equation}
where, as was already mentioned above, the superscript $\mu $
acquires the values \textquotedblleft $+$\textquotedblright\ and
\textquotedblleft $-$\textquotedblright.\,\,Therefore, for the sake
of convenience, we split the expression obtained for
$V_{Q,1,0}^{(\rm shell)}$ into $V_{Q,1,0}^{+(\rm shell)}$ and
$V_{Q,1,0}^{-(\rm shell)}$, respectively.\,\,The integrals $I_{n}$
were calculated nu\-merically.

\subsection{Inelastic electron scattering\\ giving rise to the molecule
excitation}

\label{molecula}

For an electron scattered by the molecule, the Hamiltonian of their
interaction looks like
\begin{equation}
H_{\rm int}=-eW(\mathbf{r})(c^{\dagger }+c)=-e\frac{\mathbf{d}\,
\mathbf{r}}{r^{3}}(c^{\dagger }+c),
\end{equation}%
where $c^{\dagger }$ and $c$ are the creation and annihilation,
respectively, operators for the molecule transition from the ground
state into the excited one and backward.\,\,Taking expression
(\ref{matrix}) for the scattering amplitude into account, the matrix
elements of the system should be sought in the form
\begin{equation}
W_{Q}^{(\rm dip)}=-e\int W(\mathbf{r})e^{i\mathbf{Q}\,
\mathbf{r}}d^{3}r.
\end{equation}%
As was done in the case of a nanoshell, we select the direction of
the applicate axis along the vector of transferred momentum
$\mathbf{Q}$.\,\,Changing to the spherical coordinate system and
making allowance for the series expansion (\ref{exp_expand}) of
exponential function, we can integrate over the angles $\phi $ and
$\theta $.\,\,Since $\mathbf{d}$ is directed along the vector
$\mathbf{Q}$, the component $d_{z}$ can be expressed in terms of
those two vectors, which are independent of the radial variable $r$.
As a result of the indicated operations, we obtain the following
expression for the matrix element $W_{Q}^{(\rm dip)}$:
\begin{equation}
W_{Q}^{(\rm dip)}=4\pi ei\frac{\mathbf{d}\,
\mathbf{Q}}{Q}\int\limits_{0}^{\infty }j_{l}(Qr)dr=4\pi
ei\frac{\mathbf{d}\, \mathbf{Q}}{Q^{2}}. \label{dipol}
\end{equation}%
The resonant energy, which is used in expression (\ref{doublecross})
for the double differential cross-section is the energy of the
dipole transition between the molecule le\-vels,~$E_{0}$.

\subsection{Inelastic electron scattering\\ by the system consisting of the
molecule\\ and the nanoshell}

The operator of electron interaction with the system consisting of
the molecule and the nanoshell can be written in the
form\vspace*{-1mm}
\begin{equation}
H_{\rm
int}=\sum_{m}(V_{m}^{\mu}\hat{Q}_{m}^{\mu\dagger}+V_{m}^{\mu\ast}\hat
{Q}_{m}^{\mu})+W\hat{c}^{\dagger}+W^{\ast}\hat{c},
\end{equation}
where $V_{m}^{\mu}$ is the potential of interaction between the electron and
plasmons on the nanoshell surface, and $W$ the potential of interaction
between the molecule and the electron.

For the further calculation of matrix elements, let us express the
perturbation in terms of the mixed state operators of the system,
$\hat{\psi}_{j}^{\dagger}$ and $\hat{\psi}_{j}$, which are defined
by formulas (\ref{Q}) and (\ref{c}),\vspace*{-2mm}
\[H_{\rm
int}=\sum_{j=1}^7\Biggl(\,\sum_{p=1}^3V_p^+(\mathbf{r})S_{pj}\,+\]\vspace*{-7mm}
\begin{equation}
 + \sum_{p=3}^6V_p^-(\mathbf{r})S_{pj}
+W(\mathbf{r})S_{7j}\!\Biggr)\hat{\psi}^\dagger_j+{\rm c.t.}
\end{equation}\vspace*{-3mm}

With regard for the expressions for the electron scattering at the
nanoshell, Eq.\,(\ref{nanoshell}), and the molecule,
Eq.\,(\ref{dipol}), we obtain the following formula for the matrix
elements $V_{Qn}(\Theta_{\rm sc})$ of the electron transition at its
scattering by the system consisting of the nanoshell and the
molecule:\vspace*{-1mm}
\[
V_{Qn}(\Theta_{\rm sc}) = S_{7j}W_{Q}^{(\rm dip)}
+(S_{1j}\sin\Theta_{\rm sc}\cos\phi_{\rm sc}\,+\]\vspace*{-8mm}
\[ +\,S_{2j}\sin\Theta_{\rm sc}\sin\phi_{\rm sc} +
 S_{3j}\cos\Theta_{\rm sc})
V_{Q, 1,0}^{+(\rm shell)}\, +\]\vspace*{-8mm}
\[+\,(S_{4j}\sin\Theta_{\rm sc}\cos\phi_{\rm sc}\,
+\]\vspace*{-8mm}
\begin{equation}
 +\,S_{5j}\sin\Theta_{\rm sc}\sin\phi_{\rm sc}
+S_{6j}\cos\Theta_{\rm sc})V_{Q,
  1,0}^{-(\rm shell)}
\end{equation}
where $\Theta_{\rm sc}$ and $\phi_{\rm sc}$ are the electron
scattering angles.\,\,The coefficients $S_{ij}$ can be calculated by
diagonalizing the Hamiltonian (Section 2.2).\,\,The integrals
$V_{Q,1,0}^{\mu(\rm shell)}$ are calculated independently and,
therefore, have the same form as in Section 3.1. The difference
consists is that the coefficients $S_{ij}$ include the interaction
between the molecular excitation and the plasmon modes.\,\,For
example, the coefficient $S_{7j}$ can exceed 1 very much, because
the molecule excitation \textquotedblleft borrows\textquotedblright\
the oscillator strength from the plasmon. Just this difference
increases the probability of the molecule excitation near metallic
nanoshells.

\begin{figure}%
\vskip1mm
\includegraphics[width=\column]{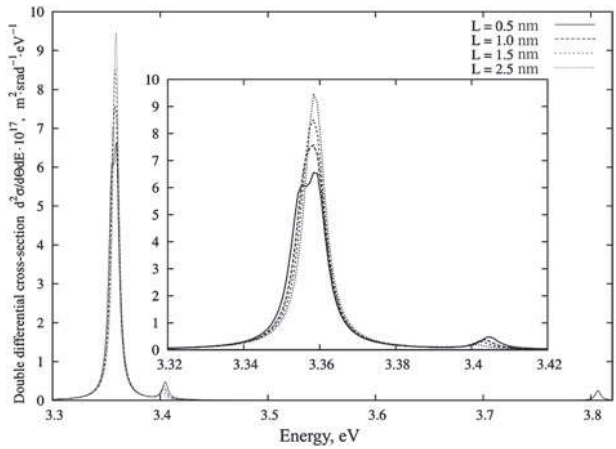}
\vskip-3mm\caption{EEL spectra for various distances $L$ between the
shell and the molecule.\,\,Electrons are scattered by the system
consisting of a silver shell with the outer radius
{$R_{2}\,{=}\,7.5$}\textrm{~nm} and the inner radius
{$R_{1}\,{=}\,4$}\textrm{~nm ,} and a molecule with the dipole
moment {$d\,{=}\,10~$}D.\,\,${E_{k}=10~}\mathrm{keV}${, $\hbar
w_{p}=11.5853$}$~\mathrm{eV}${, $\varepsilon_{0}=8.926$, and
$\hbar\omega_{0}\,{=}\,3.4$}$~\mathrm{eV} ${.} The molecule has the
orientation $\left( {\Theta\,{=}\,\pi/6,\phi\,{=}\,\pi/6}\right) $
with respect to the shell }\label{ris:Distance}
\end{figure}

\section{Results and Discussion}

The EEL spectra were calculated for a silver nanoshell, the
dielectric function of which is described by the parameters: $\hbar
\omega _{p}=11.5853~\mathrm{eV}$ and $\varepsilon _{0}=8.926$. The
inner nanoshell radius was selected to be 4\textrm{~nm}, whereas the
outer radius was varied from 7.5 to 10\textrm{~nm}.\,\,The dipole
moment of the molecule changed in the limits from 5 to 12~D.\,\,In
the majority of calculations, the molecule was considered to be at a
distance of 1\textrm{~nm} from the shell surface.\,\,The energy of
dipole transition in the molecule was $\hbar \omega
_{0}=3.4~\mathrm{eV}$.\,\,The energy of the fast electron was
10$^{4}$~eV.\,\,The transferred momentum $Q$ increases as the
scattering angle grows.\,\,Since the scattering amplitude decreases
for larger transferred momenta at least as $1/Q^{2}$, the
calculations were carried out for the angles $\Theta _{\rm sc}=0$
and $\phi _{\rm sc}=0$, i.e.\,\,the forward scattering was
considered.

While calculating the double differential scattering cross-section,
the delta-function was replaced by the Lorentzian\vspace*{-2mm}
\begin{equation}
\delta (\Delta E-E_{j})=\frac{1}{\pi }\frac{\Gamma }{(\Delta
E-E_{j})^{2}+\Gamma ^{2}},
\end{equation}\vspace*{-4mm}

\noindent where $\Gamma $ is a phenomenological parameter that
describes the line broadening for the surface plasmon.\,\,In our
calculations, we took $\Gamma =3$$\div$$ 5$~meV.

\begin{figure}%
\vskip1mm
\includegraphics[width=\column]{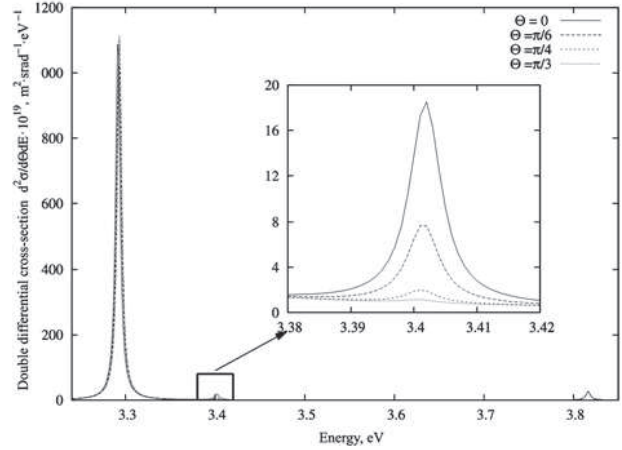}
\vskip-3mm\caption{EEL spectra for various angles $\theta$ of the
molecule orientation with respect to the nanoshell.\,\,Electrons are
scattered by the system consisting of a silver shell and a molecule
at a distance of 1\textrm{~nm}.\,\,The other parameters are the same
as in Fig.~4 }\label{ris:Theta}
\end{figure}

\begin{figure}%
\vskip3mm
\includegraphics[width=\column]{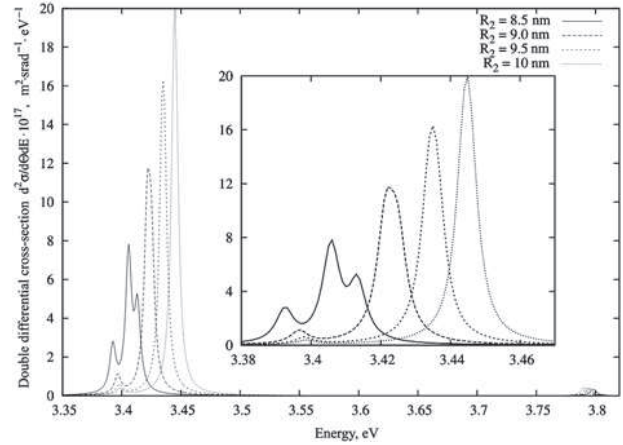}
\vskip-3mm\caption{EEL spectra for various ratios $R_{2}/R_{1}$
between the outer and inner nanoshell radii.\,\,The outer radius
$R_{2}$ was varied, the inner one was fixed, $R_{1}=4$\textrm{~nm.
}The distance between the molecule and the shell equals 1~nm.\,\,The
other parameters are the same as in Fig.~4   }\label{ris:R2}
\end{figure}

In Figs.\,\,4 to 6, three peaks are observed.\,\,The most intensive
peak is located at the oscillation frequency of a plasmon on the
outer shell surface.\,\,The peak corresponding to the molecule
excitation enhancement is located near to it.\,\,The least intensive
peak is located at the oscillation frequency of a plasmon on the
inner shell surface (at about 3.8~eV).\,\,The inset in each figure
demonstrates the scaled-up right section of the corresponding
spectrum.

The main contribution to the molecule excitation enhancement is
given by the plasmon mode on the outer nanoshell surface.\,\,The
frequency of the modes on the nanoshell surfaces depends on the
ratio between the inner and outer radii.\,\,By varying this ratio,
we can obtain the required resonant frequency of plasmons (Fig.~6).
This is the advantage of a shell in comparison with a solid
nanoparticle, on the surface of which only one mode, whose frequency
is almost independent of the particle
 radius, emerges.\,\,Figure~6
illustrates the \textquotedblleft borrowing\textquotedblright\ of
the oscillator strength by the molecule from the plasmon
mode.\,\,Under resonant conditions, this effect grows (the most
intensive peak decreases, and the peak associated with the molecule
increases). In addition, the levels of the molecule and the surface
plasmon shift, as was mentioned in Section 2.2.

In Fig.~4, the dependence of the molecule excitation probability on
the distance between the shell and the molecule is shown.\,\,As the
distance grows, the probability of molecule excitation decreases and
almost completely disappears, if the distance is approximately twice
as large as the metallic shell thickness.

The amplitude of the inelastic electron scattering by the molecule
is given by Eq.~(\ref{dipol}) and depends on the scalar product of
the transferred momentum and the dipole moment of the
transition.\,\,This dependence holds true in the presence of a
nanoshell as well.\,\,Since the forward scattering is considered,
the direction of the vector $\mathbf{Q}$ coincides with the
direction of the electron beam propagation.\,\,Hence, if the dipole
moment of the molecule transition is perpendicular to the electron
beam direction, electrons will not be scattered by the molecule.
Accordingly, the electron scattering will be maximum if the dipole
moment of the molecule transition is parallel to the beam.

If the orientation angle $\theta$ of the molecule with respect to
the nanoshell increases, whereas the orientation of the dipole
transition in the molecule remains the same, the enhancement effect
decreases (Fig.~5) and completely disappears already at a value of
$\pi/3$.\vspace*{-2mm}

\section{Conclusions}

A quantum-mechanical model of inelastic electron scat\-te\-ring by
the system consisting of an organic molecule and a spherical silver
nanoshell was de\-ve\-lo\-ped.\,\,The corresponding energy-loss
spectra were calculated and ana\-ly\-zed.\,\,The spectral analysis
showed that the main contribution to the molecular excitation
enhancement is given by the plasmon mode, which approximately
corresponds to charge oscillations on the outer shell
sur\-fa\-ce.\,\,The forward electron scattering dominates in the
spectra.

The resonant interaction between the molecular excitations and the
localized plasmons in the silver nanoshells was shown to result in a
substantial growth of the molecule excitation probability.\,\,This
enhancement of the scattering cross-section creates a possibility to
observe the excitation of molecules that form aggregates with
metallic nanoshells in an electron microscope even if they are not
observed without nanoparticles.\,\,The condition of resonance
between molecular excitations and the nanoshell plasmon mode can be
satisfied by a proper selection of the ratio of the outer and inner
nanoshell radii.

\subsubsection*{\!\!\!\!\!\!APPENDIX}

{\footnotesize The coefficients in the definition of the normal modes
for a plasmon oscillation are
\[
\Delta_{l} = \frac{l(l+1)}{R_2^{2l+1}} \left[ \varepsilon^{(3)} -
  \varepsilon_{0} \right]\! \left[ \varepsilon_{0} - \varepsilon^{(1)}
  \right] +
\]
\begin{equation}
 +\,\frac{1}{R_1^{2l+1}} \left[ \varepsilon^{(1)}l +
  \varepsilon_{0}(l+1) \right]\! \left[ \varepsilon^{(3)}(l+1) +
  \varepsilon_{0}l \right]\!,
  \end{equation}
\begin{equation}
\left\{\!\!
\begin{array}{l}
A_{lm}^{11} = -\frac{l}{\Delta_l} \left\{\! \left[ \varepsilon^{(3)}
- \varepsilon_{0}\right]\frac{l+1}{R_2^{2l+1}}\, + \right. \\[2mm]
 \left. +\,\frac{1}{R_1^{2l+1}} \left[\varepsilon^{(1)}l + \varepsilon_{0}(l+1)\right] \!\right\}\!, \\[2mm]
A_{lm}^{12} = \frac{l+1}{R_2^{2l+1}R_1^{2l+1} \Delta_l} \left(  \varepsilon^{(3)}  (l+1) +  \varepsilon^{(1)}l \right)\!,\\[2mm]
A_{lm}^{21} = \frac{l}{\Delta_l}  \left( \varepsilon^{(3)}(l+1)  + \varepsilon^{(1)}  \right)\!,\\[2mm]
A_{lm}^{22} = \frac{l+1}{\Delta_l}\left(\!l \frac{\varepsilon_{0} -
\varepsilon^{(1)}}{R_2^{2l+1}} - \frac{\varepsilon^{(3)} (l+1) +
\varepsilon_{0}l}{R_1^{2l+1}}\! \right)\!.
\end{array}
\right.
\end{equation}

The coefficients in the relation between $Q_{lm}$ and $q_{lm}$
are
\[
{B_{lm}^{\mu}}={\frac{R_2^{2l+1}-R_1^{2l+1}}{4\pi\omega
_p^2}}\,\times
\]
\begin{equation}
 \times\, \sqrt{l + \frac{l+1}{R_2^{2l+1}R_1^{2l+1}}\left(\! \frac{
\omega
    _l^{\mu}+\omega _p^2A_{lm}^{11}}{\omega _p^2A_{lm}^{12}}\!
    \right)^{\!\!2}  }.
\end{equation}

The coefficients in the formulas for the field in three nanoshell regions
are
\[
N_l^\mu = -A^{11}_{lm} - A^{21}_{lm}R_1^{2l+1}\, + \]
\begin{equation}
 +\left(\!
\frac{(\omega_l^\mu)^2
+\omega_p^2A^{11}_{lm}}{\omega_p^2A^{12}_{lm}}\!\right)
(A^{22}_{lm}R_1^{2l+1} + A^{12}_{lm}),
\end{equation}
\begin{equation}
F_l^\mu = - A^{21}_{lm} + \frac{(\omega_l^\mu)^2 +
\omega_p^2A^{11}_{lm}}{\omega_p^2A^{12}_{lm}}  A^{22}_{lm},
\end{equation}
\[
K_l^\mu = -A^{11}_{lm}R_2^{2l+1} - A^{21}_{lm}\, +\]
\begin{equation}
+ \left(\! \frac{(\omega_l^\mu)^2 +
\omega_p^2A^{11}_{lm}}{\omega_p^2A^{12}_{lm}}\!\right)  (A^{22}_{lm}
+ R_2^{2l+1}A^{12}_{lm}).
\end{equation}

}

\vspace*{-5mm}
\rezume{%
І.Ю.\,Голіней, Є.В.\,Оникієнко}{РЕЗОНАНСНЕ ПІДСИЛЕННЯ\\
ІНТЕНСИВНОСТІ ЗБУДЖЕННЯ МОЛЕКУЛИ\\ В СПЕКТРІ
  НЕПРУЖНОГО РОЗСІЯННЯ\\ ЕЛЕКТРОНІВ ЗАВДЯКИ ВЗАЄМОДІЇ\\ З ПЛАЗМОНАМИ
  МЕТАЛЕВОЇ НАНООБОЛОНКИ} {У роботі побудовано квантово-механічну модель розрахунку спектра
енергетичних втрат швидких електронів на системі, що складається з
нанооболонки та розміщеної неподалік молекули. Показано, що у
випадку резонансу між збудженням молекули та плазмонними модами
нанооболонки перетин непружного розсіяння електронів на енергії
збудження молекули значно зростає внаслідок позичання молекулярним
переходом сили осцилятора від плазмона. Вибір співвідношення
внутрішнього та зовнішнього радіусів нанооболонки створює можливість
забезпечення резонансу. Завдяки підсиленню перетину непружного
розсіяння електрона на молекулі, створюється можливість спостерігати
її перехід в електронному мікроскопі.  Побудовано й проаналізовано
залежності спектрів енергетичних втрат швидких електронів від
взаємного розташування молекули й металевої нанооболонки, відношення
внутрішнього та зовнішнього радіусів нанооболонки, кута
розсіювання.}

\end{document}